\documentclass[twocolumn,english,pra,reprint,superscriptaddress]{revtex4-2}
\usepackage[T2A,T1]{fontenc}
\usepackage[utf8]{inputenc}
\usepackage{verbatim}
\usepackage{amsmath}
\usepackage{amssymb}
\usepackage{graphicx}
\usepackage{color}
\usepackage{braket}

\usepackage{soul}
\setstcolor{red}

\makeatletter

\usepackage{graphicx} 
\graphicspath{{plots_submit/}} 

\usepackage[colorlinks,citecolor=blue,linkcolor=red,urlcolor=blue]{hyperref}

\usepackage{amsthm}
\usepackage{braket}
\makeatother
\usepackage{babel}

\begin{document}
	
	\title{Dark-state induced trapping law in single-photon emission from multiple quantum emitters}
	
	\author{Lei Qiao}
	\email{qiaolei@nus.edu.sg}
	\affiliation{Department of Physics, National University of Singapore, Singapore 117551, Singapore}
	
	\author{Jiangbin Gong}
	\email{phygj@nus.edu.sg}
	\affiliation{Department of Physics, National University of Singapore, Singapore 117551, Singapore}
	\affiliation{Center for Quantum Technologies, National University of Singapore, Singapore 117543, Singapore}	
	
	\begin{abstract}
		We study the single-photon collective dynamics in a waveguide system
		consisting of the photon channel with a finite bandwidth and an ensemble of
		quantum emitters. The size of the volume of these quantum emitters is
		ignorable when compared with the wavelength of the radiation photons. Based
		on the analytical calculations beyond the Wigner-Weisskopf and Markovian
		theories, we present exact solutions to the time evolution of the
		excited emitters with collective effects. Different from the trapping effect
		caused by photon-emitter bound states, we find that the dark states in the
		systems lead to a universal trapping behavior independent of the bosonic bath
		and the coupling strength between photons and emitters. Instead, the
		trapping is solely determined by the number of initially excited emitters
		and the total number of emitters. We demonstrate that such a trapping law
		can persist even when there are more than one type of emitters in the system.
		Our findings lead to the prediction that single-photon collective emissions
		can be strongly suppressed if the number of excited emitters is much less than the
		total number of emitters in the system.
	\end{abstract}
	
	\maketitle
	
\section{Introduction}

The coupling of quantum emitters (QEs) to a quantized radiation field can
bring about drastically different physical phenomena depending on the
specific structure of the photon environment. In free space, the dynamics of
initially excited QEs typically exhibits exponential decay. By contrast, QEs
can undergo coherent emission and reabsorption of photons in a single-mode
cavity \cite{Walls07} as a special photon environment. In particular, with
the development of new avenues in the integration of QEs with nanophotonic
structures, there are now a variety of platforms to investigate the dynamics
of QEs coupled with radiation fields with nontrivial electromagnetic
dispersions in a confined space. Examples include systems for guided
surface plasmons coupled by individual optical emitters \cite%
{Chang07,Rycenga11}, photonic nanowire with embedded quantum dots \cite%
{Lodahl15}, and superconducting transmission line coupled by superconducting
qubits \cite{Blais21}. In these systems, the tight confinement of the
propagating electromagnetic radiation leads to the enhancement of coupling
between the QEs and photons \cite{Tame13}, yielding a number of intriguing
dynamical phenomena such as persistent quantum beats \cite{Zheng13,Song21},
unidirectional emission \cite{Petersen14,Lodahl17}, single photons by
quenching the vacuum \cite{Burillo19}, and supercorrelated radiance \cite%
{Wang20}.

The interference between coherent radiation channels in an ensemble of QEs
results in collective emission \cite%
{Cummings83,Cummings86,Benivegna88,Buzek89,Buzek99}, as first illustrated by
the Dicke superradiance and subradiance \cite{Dicke54,Gross82}. Such
collective interactions between QEs and photons play an important part in
the various applications of quantum optics such as optical quantum state
storage \cite{Eisaman04,Lvovsky09,Nicolas14}, quantum communication \cite%
{Kuzmich03,Matsukevich04}, and quantum information processing \cite%
{Porras08,Pan12}. As one prominent example representing advances in
designing and probing light-matter interactions, the collective coupling of
a macroscopic number of single-molecule magnets with a microwave cavity mode
has recently been realized \cite{Eddins14}. It is equally motivating that
the large collective Lamb shift of two distant superconducting artificial
atoms has also been observed in a superconducting transmission line
terminated by a mirror \cite{Wen19}.

The new avenues in the integration of QEs with nanophotonic structures
stimulate the investigation of physics of photon-QE interactions in
one-dimensional waveguide settings that are engineered to have nontrivial
dispersion relations with band edges and band gaps \cite%
{Lombardo14,Calajo16,Song20,Burillo17,Qiao17,Qiao19}. Near band edges or
band gaps of the photonic dispersion relation, the group velocity of the
propagating photons is greatly reduced or even completely prohibited,
triggering new possibilities. It has been demonstrated that the spontaneous
emission of an excited atom coupled to the band edge of a photonic crystal
reveals nonexponential decay dynamics, with a finite non-decaying excitation
fraction exhibiting oscillatory behaviors \cite{John94,Kofman94,Sakoda01}.
This population trapping is due to the presence of localized atom-field
bound states with energies outside the band of scattering modes \cite{John90}%
. When it comes to many QEs, the non-decaying fraction of QEs can be
attributed to two different trapping mechanisms. One comes from the
existence of photon-QE bound states, the other arises from that of dark
states with energies equal to the transition frequencies of the QEs \cite%
{Qiao19a,Qiao20}. In an ensemble of QEs confined to a small volume compared
to the radiated wavelengths, it has recently been pointed out that the
emission dynamics contributed by dark states will obey the $(1-1/M)$ trapping law ($M$ is the total number of QEs)
if only one of the QEs is excited initially \cite{Qiao19a,Qiao20}. It was also
previously shown that this kind of population-trapping law is robust in
different QE systems.

In this paper, we focus on a more general situation in the single-photon
regime, where the initial state, though in the single-photon Hilbert
subspace, involving a superposition of excitations from different QEs.
Loosely speaking, the initial excitation involves more than one QEs. We
investigate the ensuing cooperative dynamics based on an analytical analysis
beyond the Wigner-Weisskopf approximations and Markovian approximations. A
new excitation trapping law is identified. The found trapping law behavior
does not depend on the specific light-field environment or the coupling
strength between QEs and photons. As one direct application of our finding,
one can predict that if the total number of QEs is much greater than the
number of excited QEs in the initial state, the collective spontaneous
emission is strongly suppressed. The trapping properties of more than one
types of QEs are also explored and similar trapping law is found to persist under certain conditions. 
Note that throughout the paper, the QEs are assumed to be placed much closer than the
wavelength of the radiation photons and thus the QEs are effectively coupled
to the radiation field without retardation effects.

This paper is organized as follows. In Sec.~II, we introduce our model
consisting of an assembly of QEs and a coupled-resonator waveguide. In
Sec.~III, we investigate the single-photon collective dynamics in the
presence of one type of QEs. In Sec. IV, the time evolution of excited QEs
is also analyzed with different types of QEs participating in the dynamics. Finally, we summarize the
results and give our conclusions and discussions in Sec. V.

\section{Model}

We consider a system consisting of a one-dimensional array of tunnel-coupled
resonators. One of the resonators is also directly coupled with different
types of two-level QEs. The $j$th QE of type $i$ is assumed to have excited
state $\left\vert e_{j}^{i}\right\rangle $ and ground state $\left\vert
g_{j}^{i}\right\rangle $, separated in energy by frequency $\Omega _{i}$ (we
set $\hbar =1$ throughout). Denoting $a_{x}$ ($a_{x}^{\dag }$) as the
bosonic annihilation (creation) operator for a photon at site $x$, the
tight-binding Hamiltonian of the resonator-photon system can be modeled as%
\begin{align}
H& =\sum_{x}\omega _{c}a_{x}^{\dag }a_{x}+\sum_{x}J\left( a_{x+1}^{\dag
}a_{x}+a_{x}^{\dag }a_{x+1}\right)   \notag \\
& +\sum_{i}\sum_{j}\Omega _{i}\left\vert e_{j}^{i}\right\rangle \left\langle
e_{j}^{i}\right\vert   \notag \\
& +\sum_{i}\sum_{j}V_{i}\left( \sigma _{j}^{i+}a_{x_{0}}+\sigma
_{j}^{i-}a_{x_{0}}^{\dag }\right) \text{,}  \label{Hamiltonian}
\end{align}%
where $\omega _{c}$ is the resonance frequency of each resonator. $J$
represents the hopping energy of photons between two neighbouring lattices.
Here, $\sigma _{j}^{i+}=\left\vert e_{j}^{i}\right\rangle \left\langle
g_{j}^{i}\right\vert $ ($\sigma _{j}^{i-}=\left\vert g_{j}^{i}\right\rangle
\left\langle e_{j}^{i}\right\vert $) is the raising (lowering) operator
acting on the $j$th QE of type $i$. $V_{i}$ is the coupling strength between
the waveguide mode at resonator $x_{0}$ and type-$i$ QEs. For convenience,
we further assume that the lattice constant $a=1$ throughout. Such
coupled-resonator setups have been realized in different platforms, such as
the coupled superconducting cavities \cite%
{Fitzpatrick17,Sundaresan19,Ferreira21} and the coupled nanocavities in
photonic crystals \cite{Notomi08}. The typical values for the coupling
strength $V_{i}$ and hopping energy $J$ go up to a few hundred MHz in these
experiments, whereas the frequency $\Omega _{i}$ can be controllable within
a few GHz. The resonator dissipative rate $\gamma _{c}$ and the emitter
dissipative rate $\gamma _{e}$ are in the kHz regime and are thus much
smaller than $V_{i}$, $J$ and $\Omega _{i}$ \cite{Meher22}. This being the
case, the system's dissipation can be safely neglected in our theoretical
considerations below.

The first two terms in Eq. (\ref{Hamiltonian}) describe the free photon
Hamiltonian and can be diagonalized by introducing the Fourier transform%
\begin{equation}
a_{k}=\frac{1}{\sqrt{N}}\sum_{x}e^{-ikx}a_{x}\text{,}
\end{equation}%
where $k$ is the wave number within the first Brillouin zone and $k\in
\lbrack -\pi ,\pi ]$, which becomes continuous in the the limit of $%
N\rightarrow \infty $. In this $k$-representation, the free photon
Hamiltonian becomes $\sum_{k}\omega _{k}a_{k}^{\dag }a_{k}$ with the
dispersion $\omega _{k}=\omega _{c}+2J\cos \left( k\right) $. This mode
frequency vs $k$ forms a scattering band with $\omega _{c}$ being the band
center with bandwidth $4J$ ($J>0$). Such structured modes support photons to
transport in the waveguide with the group velocity $v_{g}(k)=$ $-2J\sin
\left( k\right) $, which reaches its extreme values at the center of band
and gets to zero at the two band edges. Still using the $k$-representation,
the Hamiltonian in Eq.~(\ref{Hamiltonian}) can be rewritten as%
\begin{equation}
H=\sum_{k}\omega _{k}a_{k}^{\dag }a_{k}+\sum_{i}\sum_{j}\Omega
_{i}\left\vert e_{j}^{i}\right\rangle \left\langle e_{j}^{i}\right\vert
+H_{I},
\end{equation}%
with%
\begin{equation}
H_{I}=\sum_{i}\sum_{j,k}\frac{V_{i}}{\sqrt{N}}\left( \sigma
_{j}^{i+}e^{ikx_{0}}a_{k}+\sigma _{j}^{i-}e^{-ikx_{0}}a_{k}^{\dag }\right) 
\text{.}
\end{equation}%
This expression of the system Hamiltonian indicates clearly that the QEs are
coupled to a finite-width energy band of waveguide modes. From now on, for
simplicity of calculation, $x_{0}$ is set to be the zero point of the $x$
axis. For the case of having only one QE, the spontaneous emission of the
sole QE will be much suppressed due to the population trapping effect
arising from bound states, if the QE's frequency is outside the waveguide
energy band \cite{Burillo17}. On the contrary, when the transition frequency
of the QE lies inside the band and far away from the upper and lower edges,
the excited QE will undergo an exponential decay if the coupling strength $%
V\ll 2J$ while it will exhibit stable Rabi oscillation for sufficiently long
time if the coupling strength $V\gtrsim 2J$ \cite{Lombardo14}.

\section{Dynamics and trapping law with one type of QEs}

We first investigate the situation where the waveguide system hosts only one
type of QEs. This configuration is also known as one of the general
Fano-Anderson models. In this case, there are two photon-QE bound states
with nonzero field amplitudes. One bound state's energy is above the
scattering band and the other bound state's energy is below the bottom of
the band \cite{Qiao19a}. The dynamics with only one QE being initially
excited among $M$ QEs has been investigated and a universal trapping law,
has been found in the population dynamics \cite{Qiao20}, namely, at long
time the population will be trapped at $(1-1/M)^2$. Here, we explore the
situation that multiple QEs are initially excited, where the initial state
is now an entangled state involving multiple QEs and quantum interference
between different deexcitation pathways may lead to interesting physics.

\begin{figure}[tbp]
\centering
\includegraphics[width=0.9\columnwidth]{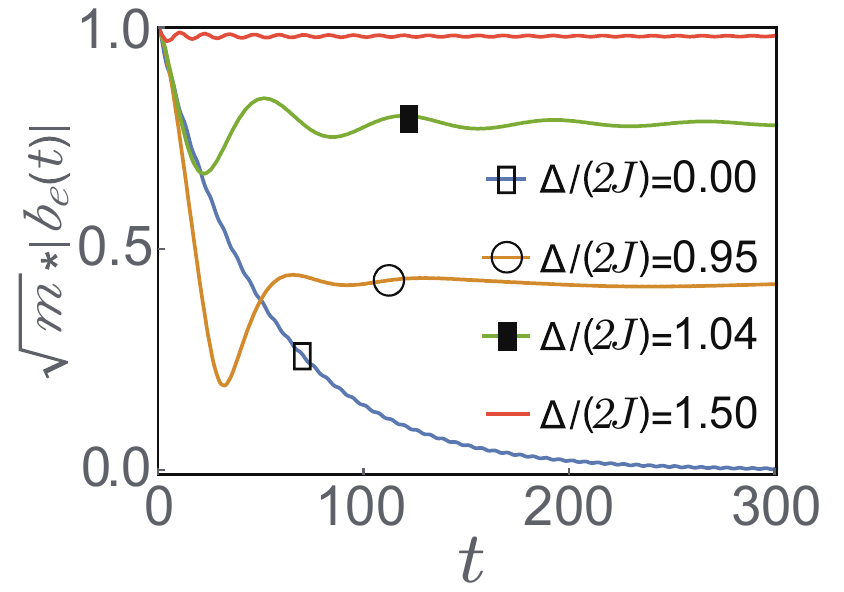}
\caption{Time evolution of  the magnitude of $b_{e}(t)$, the amplitude on the excited QEs with 
$m=3$. The time is in units of $1/(2J)$. Other parameters are $V/(2J)=0.08$
and $M=3$. Here and in all other figures, the plotted results are computed directly from our analytical results that have been also confirmed by numerical simulations based on the time-dependent 
Schr\"{o}dinger equation.}
\label{Figure1}
\end{figure}

To specifically and theoretically investigate the spontaneous emission
dynamics, we start from the time-dependent Schr\"{o}dinger equation%
\begin{equation}
i\frac{\partial }{\partial t}\left\vert \psi \left( t\right) \right\rangle
=H\left\vert \psi \left( t\right) \right\rangle \text{.}  \label{Schr}
\end{equation}%
The time-evolving state $\left\vert \psi \left( t\right) \right\rangle $ at
time $t$ can be written as $\left\vert \psi \left( t\right) \right\rangle
=\sum_{j}b_{j}(t)\left\vert e_{j},0\right\rangle +\sum_{k}C_{k}(t)\left\vert
g,1_{k}\right\rangle $, where $b_{j}(t)$ ($j=1,2,...,M$) is the excitation
amplitude for the $j$th QE in this sole type of QEs and $C_{k}(t)$ is the
amplitude for the waveguide mode with wavenumber $k$. Applying Eq.~(\ref%
{Schr}), one obtains the following dynamical equations for the amplitudes%
\begin{equation}
i\frac{\partial b_{j}\left( t\right) }{\partial t}=\Omega b_{j}\left(
t\right) +\sum_{k}\frac{V}{\sqrt{N}}C_{k}\left( t\right) \text{,}
\label{Dyeq1}
\end{equation}%
\begin{equation}
i\frac{\partial C_{k}\left( t\right) }{\partial t}=\omega _{k}C_{k}\left(
t\right) +\sum_{j}\frac{V}{\sqrt{N}}b_{j}\left( t\right) \text{.}
\label{Dyeq2}
\end{equation}%
To analytically solve these coupled dynamical equations, one may make use of
the well-known Wigner-Weisskopf theory or Markovian theory by neglecting the
contributions from any possible bound states. These two treatments can work
well in the presence of one QE with the conditions $|\Omega -\omega _{c}|$ $%
\ll 2J$ and $V\ll 2J$, under which the bound-state trapping regime can be
neglected \cite{Qiao19a}. However, such approximate treatment would not be
able to capture a potential population trapping effect as mentioned above.
To capture the impact of multiple QEs on population trapping, we must go
beyond these approximations. To that end we take a Laplace transform for
Eqs. (\ref{Dyeq1}) and (\ref{Dyeq2}) with $\tilde{b}_{j}(s)=\int_{0}^{\infty
}b_{j}(t)e^{-st}dt$ and $\tilde{C}_{k}(s)=\int_{0}^{\infty
}C_{k}(t)e^{-st}dt $. This yields%
\begin{equation}
i\left[ -b_{j}\left( 0\right) +s\tilde{b}_{j}\left( s\right) \right] =\Omega 
\tilde{b}_{j}\left( s\right) +\sum_{k}\frac{V}{\sqrt{N}}\tilde{C}_{k}\left(
s\right) \text{,}
\end{equation}%
\begin{equation}
i\left[ -C_{k}\left( 0\right) +s\tilde{C}_{k}\left( s\right) \right] =\omega
_{k}\tilde{C}_{k}\left( s\right) +\sum_{j}\frac{V}{\sqrt{N}}\tilde{b}%
_{j}\left( s\right) \text{.}
\end{equation}

Without loss of generality, we denote the initial excited QEs by index $j_{n}
$, with $j_{n}$ going from $j_{1}$, $j_{2}$,..., to $j_{m}$ if there are
initially $m$ QEs excited. All other QEs are in their ground states. Hence,
the initial conditions in terms of the initial quantum amplitudes are: $%
b_{j_{1}}(0)=...=b_{j_{m}}(0)=1/\sqrt{m}$ ($m\leqslant M$), $b_{j}(0)=0$ ($%
j\neq j_{n}$), and $C_{k}(0)=0$. After some algebra, we obtain the
expression of $\tilde{b}_{j_{1}}(s)=...=\tilde{b}_{j_{m}}(s)$ $\equiv $ $%
\tilde{b}_{e}(s)$ with $\tilde{b}_{e}(s)$ being 
\begin{equation}
\tilde{b}_{e}\left( s\right) =i\frac{is-\Omega -\left( M-m\right)
V^{2}F\left( s\right) }{\sqrt{m}\left( is-\Omega \right) \left[ is-\Omega
-MV^{2}F\left( s\right) \right] }\text{,}
\end{equation}%
where $F(s)=(1/N)\sum_{k}1/(is-\omega _{k})$. The time-evolving amplitudes
for excited QEs can then be derived by use of the inverse Laplace transform $%
b_{j_{n}}(t)=(1/2\pi i)\int_{\sigma -i\infty }^{\sigma +i\infty }\tilde{b}%
_{j_{n}}(s)e^{st}ds$, with real number $\sigma $ being sufficiently large so
that all the poles are on its left side. Note all the initially excited QEs
have the same time-dependent amplitudes due to the chosen initial state. To
calculate the integral here, the analytic properties of $\tilde{b}_{j_{n}}(s)
$ are considered in the whole complex plane except a branch cut from $%
-i(2J+\omega _{c})$ to $i(2J-\omega _{c})$ along the imaginary axis. By
using the residue theorem \cite{Riley06}, we arrive at the exact expressions
for $b_{j_{1}}(t)=...=b_{j_{m}}(t)$ $\equiv $ $b_{e}(t)$ with $b_{e}(t)$
being%
\begin{align}
b_{e}(t)& =\sum_{n}\left. \frac{s+i\Omega +i\left( M-m\right) V^{2}F\left(
s\right) }{\sqrt{m}\left[ G_{1}\left( s\right) \right] ^{\prime }}%
e^{st}\right\vert _{s=\varepsilon _{n}}  \notag \\
& +\int_{-1}^{1}\frac{4\sqrt{m}V^{2}J^{2}\sqrt{1-y^{2}}e^{i2Jyt}}{L\left(
y\right) +\pi M^{2}V^{4}}dy\text{,}  \label{bet}
\end{align}%
where 
\begin{equation}
G_{1}(s)=(s+i\Omega )G_{0}(s)
\end{equation}%
with 
\begin{equation}
G_{0}(s)=s+i\Omega +iMV^{2}F(s)  \label{G0}
\end{equation}%
and $L(y)$ is defined as 
\begin{equation}
L(y)=4\pi J^{2}(1-y^{2})(2Jy+\Omega )^{2}\text{.}
\end{equation}%
Here, $[G_{1}\left( s\right) ]^{\prime }$ represents the derivative of $%
G_{1}(s)$ with respect to $s$. $\varepsilon _{n}$ is the roots of the
equation $G_{1}(s)=0$. All these roots can be divided into two kinds. One is
the solutions to the equation $G_{0}(s)=0$. This kind of roots are pure
imaginary numbers, with their imaginary parts corresponding to minus
eigenenergies of localized photon-QE bound states \cite{Qiao19a}. The other
additional root is $s=-i\Omega $ and this root corresponds to the energy of
dark states. In fact, according to the analysis using a complete basis
expansion based on Green's function method \cite{Qiao19}, the terms with $s$
being the solutions to the equation $G_{0}(s)=0$ in Eq.~(\ref{G0}) comes
from the contribution of system's photon-QE bound states.
 The second line in Eq.~(\ref{bet}) (which becomes zero
in the limit of $t\rightarrow \infty $) arises from the contribution of
system's scattering states. When the number of initial excited QEs is equal
to the total number of QEs, i.e., $m=M$, one can obtain $b_{e}(\infty
)=\sum_{n}e^{st}/\{\sqrt{M}[G_{0}(s)]^{\prime }\}|_{t\rightarrow \infty ,%
\text{ }s=\varepsilon _{n}^{\prime }}$ where $\varepsilon _{n}^{\prime }$ is
solutions to the equation $G_{0}(s)=0$. The purely imaginary roots $%
\varepsilon _{n}^{\prime }$ reveal that the populations on the excited QEs
are fractionally trapped when $t\rightarrow \infty $.

In Fig.~\ref{Figure1}, we plot the time dependence of $b_{e}(t)$ with $m=M=3$
and different detuning $\Delta =\Omega -\omega _{c}$. It can be seen that a
larger fraction of the population is trapped at long time as the transition
frequency $\Omega $ shifts away from the frequency $\omega _{c}$ of single
resonator, and the spontaneous emission is almost totally suppressed when $%
\Omega $ is far away from the energy band. Exploring many examples, it is
found that only under the condition $\sqrt{M}V\ll 2J$ and $|\Omega -\omega
_{c}|$ $<J$ with which the first term in Eq. (\ref{bet}) can be ignored, the
emission of QEs can be nearly complete and then display a basically
exponential decay, with a slowly changing radiation rate as $\Omega $ varies
from $\omega _{c}\pm J$ to $\omega _{c}$. For such a case, $\left\vert
b_{e}(t)\right\vert ^{2}$ can be approximately calculated as $\left\vert
b_{e}(t)\right\vert ^{2}\approx (1/M)e^{-\Gamma _{s}\left( \Delta \right) t}$%
, with a decay rate $\Gamma _{s}(\Delta )=2\pi MV^{2}D(\Delta )$, where $%
D(\Delta )$ is the density of states for the free-photon Hamiltonian. That is,
only for such situations the spontaneous emission dynamics is best
approximated by the Wigner-Weisskopf and Markovian approximate theory \cite%
{Scully97,Qiao22}. Note also that under the condition of $\Delta =0$, $%
D(\Delta )$ gets its extremum, and $\Gamma _{s}(0)=MV^{2}/J$, which is $M$
times the radiation rate for the case of only one QE. This is precisely what
a standard superradiance theory predicts.

\begin{figure}[tbp]
\centering
\includegraphics[width=0.9\columnwidth]{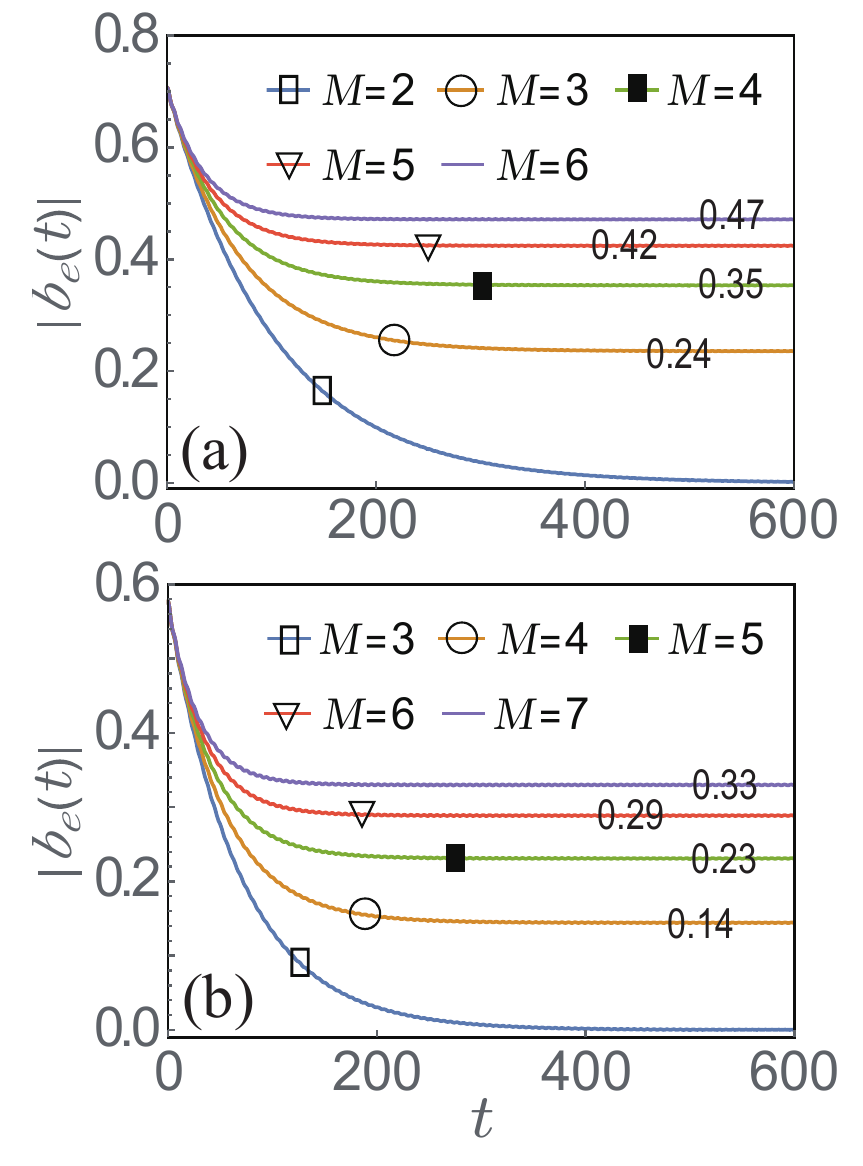}
\caption{Time evolution of the magnitude of the excited-state amplitude $b_{e}(t)$ with
different QE number $M$ for (a) $m=2$ and (b) $m=3$. Other parameters are $%
\Delta/(2J)=0$ and $V/(2J)=0.07$. Time is in units of $1/(2J)$.}
\label{Figure2}
\end{figure}


Consider next what happens if the number of initially excited QEs is less
than the total number of QEs, i.e., $m<M$. With the number $M\geqslant 2$,
there are not only nonlocalized scattering states and localized photon-QE
bound states, but also degenerate dark states with energy $E=\Omega $ \cite%
{Qiao19a,Qiao20,Sun03}. These dark states have a specific property, namely,
due to collective interference effects, these dark states allow only the QEs to be excited and so
excitation amplitudes on the photon field modes are all zero. Therefore, now both such dark states and photon-QE bound states play a role in
spontaneous emission dynamics. For the condition $\sqrt{M}V\ll 2J$ and $%
|\Omega -\omega _{c}|$ $<J$, with which the contributions from the photon-QE
bound states are much smaller than that of the dark states, then the final
values of $|b_{e}(t)|$ are found to depend only on the number $m$ of initial
excited QEs and the total number $M$ of QEs. Specifically, for sufficiently
long time (the second line of Eq.~(\ref{bet}) can be dropped due to the
highly oscillatory integral there) and upon neglecting the contributions
from the roots of $G_{0}(s)=0$ that represent contributions from the
photon-QE bound states, Eq.~(\ref{bet}) reduces to 
\begin{eqnarray}
b_{e}(t) &\approx &\sum_{n}\left. \frac{s+i\Omega +i\left( M-m\right)
V^{2}F\left( s\right) }{\sqrt{m}\left[ G_{1}\left( s\right) \right] ^{\prime
}}e^{st}\right\vert _{s=\varepsilon _{n}}  \notag \\
&\approx &\left. \frac{s+i\Omega +i\left( M-m\right) V^{2}F\left( s\right) }{%
\sqrt{m}\left[ G_{1}\left( s\right) \right] ^{\prime }}e^{st}\right\vert
_{s=-i\Omega }  \notag \\
&=&\frac{M-m}{\sqrt{m}M}e^{-i\Omega t}  \label{TrappingLaw}
\end{eqnarray}%
This is one main result of this work.

In Fig.~\ref{Figure2} (a), we plot our purely theoretical result of $%
|b_{e}(t)|$ assuming that two QEs are initially excited with different
values of $M$, the total number of QEs in the system. As time lapsed is long
enough, the final value of $|b_{e}(t)|$ during the emission dynamics is
stabilized at $(M-2)/(\sqrt{2}M)$. Similarly, for the case where three QEs
are initially excited, the amplitudes $b_{j_{1}}(t)=b_{j_{2}}(t)=b_{j_{3}}(t)
$ $\equiv $ $b_{e}(t)$ at long time are found to stabilize at $|b_{e}(\infty
)|$ $=(M-3)/(\sqrt{3}M)$ without further decay, as shown in Fig.~\ref{Figure2}
(b).  Note again that the plotted results are computed directly from our analytical results derived above and have been
also confirmed by our numerical results based on the time-dependent Schr\"{o}dinger equation.
These specific results hence have  clearly illustrated our main theoretical prediction. For $%
m=1$, $\left\vert b_{e}(\infty )\right\vert $ comes back to the previous
result already studied in the context of vacuum photonic bath, photonic
crystal and coupled-resonator waveguide \cite{Qiao20}. It is worth noting
that Eq.~(\ref{TrappingLaw}) can also include the result with $m=M$. No
trapping happens for this case and the initial energy of all QEs will be
fully released, as also illustrated in Fig.~\ref{Figure2} (a) and (b).
Physically, this is because the initial excited states with $M=m$ are
orthogonal to the dark states and as such, the presence of dark states is
unable to saturate the spontaneous decay. From the view of prolonging the
lifetime of QEs, we can see that under the condition $m\ll M$, our theory
above predicts that the spontaneous emission of the QEs will be greatly
suppressed.

\section{Dynamics and trapping law with two types of QEs}

We now investigate the properties of dynamics in the waveguide system with
two different types of QEs indexed by $A$ and $B$. Unlike the case with one
sole type of QEs where there are always two photon-QE bound states, the
energy-level structure of the system with two type of QEs can undergo
certain transitions when some system parameters change \cite{Qiao19a}, such
as the QE numbers $M_{A}$ and $M_{B}$ or the coupling strengths $V_{A}$ and $%
V_{B}$. When overall only one QE is initially excited (without loss of
generality, assuming that the excited QE belongs to type $A$), then
previously it was found that the asymptotic value of the magnitude of the
quantum amplitude of the QE is given by $1-1/M_{A}$ \cite{Qiao19a}.
Encouraged by our results from the previous section, here we wish to examine
if there is some similar trapping law if more than one QEs, but still
belonging to the same type, are initially excited. One main complication in
answering this question is that the two types of QEs can interact strongly
with each other through the waveguide system. As such, to observe an
interesting trapping law it is necessary to find under what theoretical
conditions the decay dynamics can still exhibit some trapping law. A
violation of the trapping law sought after here will give us strong
indication of the interplay between different types of QEs.

Let us now proceed with our theoretical framework. In the single excitation
subspace, the time-evolving state at time $t$ can be written as%
\begin{equation}
\left\vert \varphi \left( t\right) \right\rangle
=\sum_{i}\sum_{j}b_{j}^{i}\left( t\right) \left\vert
e_{j}^{i},0\right\rangle +\sum_{k}C_{k}\left( t\right) \left\vert
g,1_{k}\right\rangle \text{,}
\end{equation}%
where $b_{j}^{i}(t)$ ($i=A,B$) is the excitation amplitude of the system's
state for the $j$th QE of type $i$, with no photon in the waveguide. $%
C_{k}(t)$ is the amplitude for the state that all QEs are in their ground
states and there is a photon with wavenumber $k$. Plugging $\left\vert
\varphi (t)\right\rangle $ into the Schr\"{o}dinger equation $i\partial
\left\vert \varphi \left( t\right) \right\rangle /\partial t=H\left\vert
\varphi \left( t\right) \right\rangle $, one can obtain the following
coupled equations for $b_{j}^{i}\left( t\right) $ and $C_{k}\left( t\right) $%
\begin{equation}
i\frac{\partial }{\partial t}b_{j}^{i}\left( t\right) =\Omega
_{i}b_{j}^{i}\left( t\right) +\sum_{k}\frac{V_{i}}{\sqrt{N}}C_{k}\left(
t\right) \text{,}  \label{2Dyeq1}
\end{equation}%
\begin{equation}
i\frac{\partial }{\partial t}C_{k}\left( t\right) =\omega _{k}C_{k}\left(
t\right) +\sum_{i}\sum_{j}\frac{V_{i}}{\sqrt{N}}b_{j}^{i}\left( t\right) 
\text{.}  \label{2Dyeq2}
\end{equation}%
Similar to the steps in the case of one type of QEs, one can take a Laplace
transform for Eqs. (\ref{2Dyeq1}) and (\ref{2Dyeq2}) with $\tilde{b}%
_{j}^{i}(s)=\int_{0}^{\infty }b_{j}^{i}(t)e^{-st}dt$ and $\tilde{C}%
_{k}(s)=\int_{0}^{\infty }C_{k}(t)e^{-st}dt$, which leads to%
\begin{equation}
i\left[ -b_{j}^{i}\left( 0\right) +s\tilde{b}_{j}^{i}\left( s\right) \right]
=\Omega _{i}\tilde{b}_{j}^{i}\left( s\right) +\sum_{k}\frac{V_{i}}{\sqrt{N}}%
\tilde{C}_{k}\left( s\right) \text{,}  \label{Adeq1}
\end{equation}%
\begin{equation}
i\left[ -C_{k}\left( 0\right) +s\tilde{C}_{k}\left( s\right) \right] =\omega
_{k}\tilde{C}_{k}\left( s\right) +\sum_{i}\sum_{j}\frac{V_{i}}{\sqrt{N}}%
\tilde{b}_{j}^{i}\left( s\right) \text{.}  \label{Adeq2}
\end{equation}%
Let us now assume that initially only one type of QEs are excited and
without loss of generality, the excited QEs denoted by index $j_{n}$ are
assumed to be of type $A$. We further assume that in total $m_{A}$ QEs of
type $A$ are initially excited and all other QEs are in their ground state,
with amplitudes $b_{j_{1}}^{A}(0)=...=b_{j_{m_{A}}}^{A}(0)=1/\sqrt{m_{A}}$ ($%
m_{A}\leqslant M_{A}$), $b_{j}^{A}(0)=0$ ($j\neq j_{n}$), $b_{j}^{B}(0)=0$
and $C_{k}(0)=0$. After some necessary algebraic operations with Eq.~(\ref%
{Adeq1}) and (\ref{Adeq2}), one can arrive at the expression of $\tilde{b}%
_{j_{1}}^{A}(s)=...=\tilde{b}_{j_{m_{A}}}^{A}(s)$ $\equiv $ $\tilde{b}%
_{e}^{A}(s)$ with $\tilde{b}_{e}^{A}(s)$ being%
\begin{align}
\tilde{b}_{e}^{A}\left( s\right) & =i\frac{K_{A}\left( s\right) K_{B}\left(
s\right) +m_{A}V_{A}^{2}F\left( s\right) K_{B}\left( s\right) }{\sqrt{m_{A}}%
\left( is-\Omega _{A}\right) Y\left( s\right) }  \notag \\
& -i\frac{\left( M_{A}-m_{A}\right) M_{B}\left[ V_{A}V_{B}F\left( s\right) %
\right] ^{2}}{\sqrt{m_{A}}\left( is-\Omega _{A}\right) Y\left( s\right) }
\end{align}%
where 
\begin{equation}
K_{i}(s)=is-\Omega _{i}-M_{i}V_{i}^{2}F(s)\text{,}
\end{equation}%
and$\ $%
\begin{equation}
Y(s)=K_{A}(s)K_{B}(s)-M_{A}M_{B}[V_{A}V_{B}F(s)]^{2}\text{.}
\end{equation}%
Just like what was done in the previous section, the time dependence of the
excited QEs can be calculated by the inverse Laplace transform. 
Because $\tilde{b}_{j_{n}}^{A}(s)$ is an analytic function in the whole
complex plane except a branch cut from $-i(2J+\omega _{c})$ to $i(2J-\omega
_{c})$ along the imaginary axis, the exact expressions of the excited
amplitudes $b_{j_{1}}^{A}(t)=...=b_{j_{m_{A}}}^{A}(t)$ $\equiv $ $%
b_{e}^{A}(t)$ can be acquired by using the residue theorem \cite{Riley06}
and $b_{e}^{A}(t)$ is obtained as%
\begin{align}
b_{e}^{A}(t)& =\left. \frac{\left( M_{A}-m_{A}\right) }{\sqrt{m_{A}}M_{A}}%
e^{st}\right\vert _{s=-i\Omega _{A}}  \notag \\
& -\sum_{n}\left. \frac{\left( s+i\Omega _{B}\right) m_{A}V_{A}^{2}F\left(
s\right) }{\sqrt{m_{A}}\left( is-\Omega _{A}\right) \left[ G_{2}\left(
s\right) \right] ^{\prime }}e^{st}\right\vert _{s=\tilde{\varepsilon}_{n}} 
\notag \\
& +\sum_{\alpha =\pm 1}\int_{-1}^{1}\frac{J\left( 2Jy+\Omega _{B}\right)
m_{A}V_{A}^{2}f\left( y\right) e^{i2Jyt}}{\pi \sqrt{m_{A}}\left( 2Jy+\Omega
_{A}\right) Z_{\alpha }\left( y\right) }dy  \label{Dtrapping}
\end{align}%
where 
\begin{eqnarray}
G_{2}(s) &=&(is-\Omega _{A})(is-\Omega _{B})-(is-\Omega
_{A})M_{B}V_{B}^{2}F(s)  \notag \\
&&-(is-\Omega _{B})M_{A}V_{A}^{2}F(s)
\end{eqnarray}%
and $\tilde{\varepsilon}_{n}$ is the roots of the equation $G_{2}(s)=0$. We
stress that $G_{2}(-iE)=0$ can be used to determine the eigenenergies $E$ of
localized photon-QE bound states \cite{Qiao19a}. Because $G_{2}(s)$ involves
physical properties of both types of QEs, in general one anticipates that
photon-QE bound states here can lead to rather complicated population
trapping behavior \cite{John94}. As to the third term from the contribution
of the scattering states in Eq.~(\ref{Dtrapping}), it contains functions $%
f(y)$ and $Z_{\pm }(y)$ defined as $f(y)=1/(2J\sqrt{1-y^{2}})$ and $Z_{\pm
}(y)=(2Jy+\Omega _{A})(2Jy+\Omega _{B})\pm i[(2Jy+\Omega
_{A})M_{B}V_{B}^{2}+(2Jy+\Omega _{B})M_{A}V_{A}^{2}]f(y)$. As expected, we
also see a highly oscillatory factor $e^{i2Jyt}$ in the long time limit,
thus killing the third term for long time dynamics.

\begin{figure}[tbp]
\centering
\includegraphics[width=0.9\columnwidth]{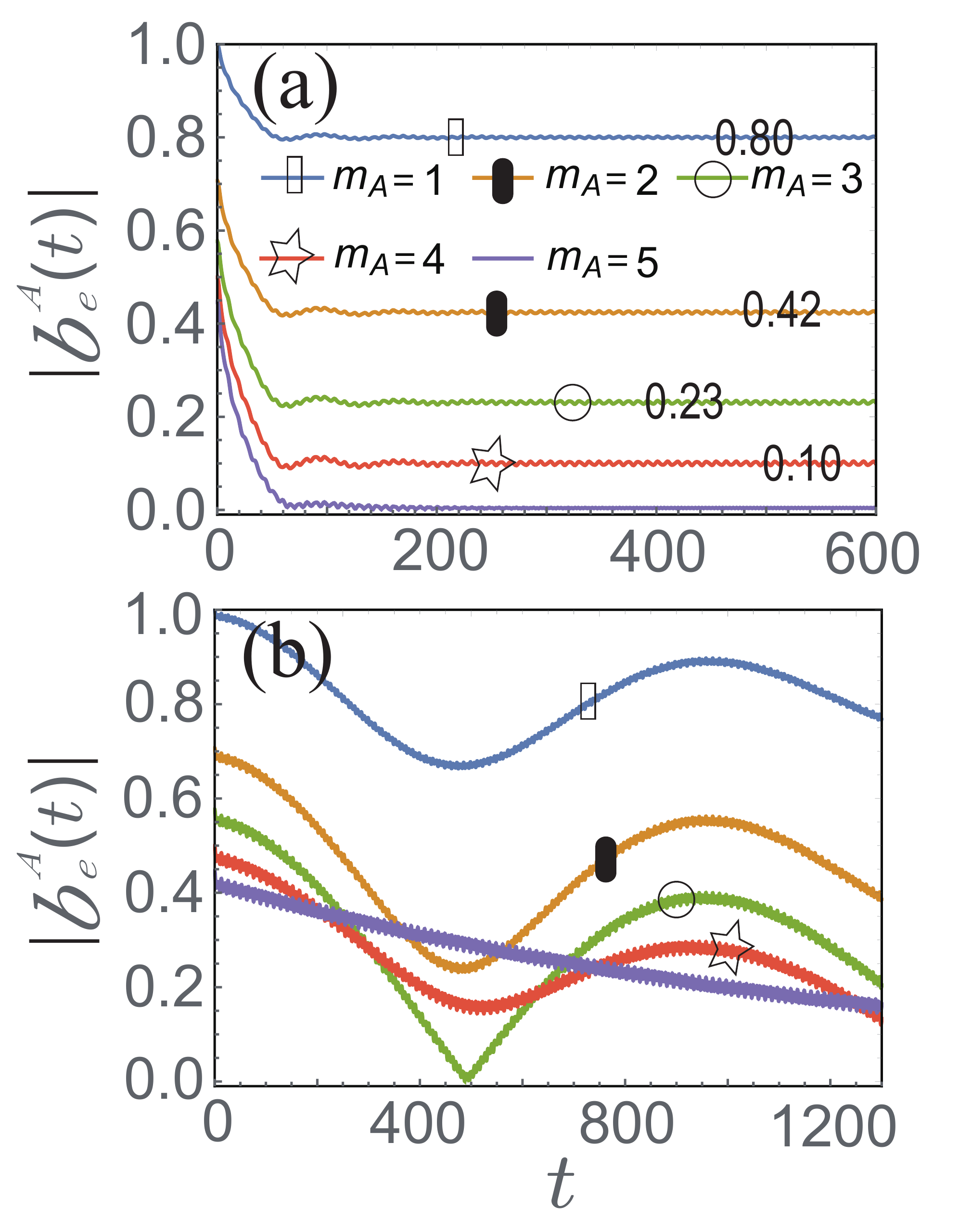}
\caption{Time evolution of the magnitude of the excited-state amplitude $b_{e}^{A}(t)$ with
different number $m_{A}$ of initially excited QEs in type $A$ for (a) $%
V_{B}/(2J)=0.1$ and (b) $V_{B}/(2J)=0.6$. Other parameters are: $\Delta
_{A}/(2J)=0.3$, $\Delta _{B}/(2J)=0.2$, $V_{A}/(2J)=0.1$, $M_{A}=5$ and $%
M_{B}=2$. The time is in units of $1/(2J)$.}
\label{Figure3}
\end{figure}

Despite the complicated contributions from the photon-QE bound states
involving two types of QEs, what we learned from the previous section is
that there are still a wide parameter regime where we may focus on the
contributions of the dark state only with the bound-state contributions
being negligible. That is, if the magnitude of $(s+i\Omega
_{B})m_{A}V_{A}^{2}F(s)/\{\sqrt{m_{A}}(is-\Omega _{A})[G_{2}(s)]^{\prime
}\}|_{s=\tilde{\varepsilon}_{n}}$ is sufficiently small, which may be
satisfied, e.g., under the condition $V_{A}\ll 2J$, the asymptotic amplitude 
$b_{e}^{A}(t)$ can be easily identified as well, namely, 
\begin{equation}
\left\vert b_{e}^{A}(\infty )\right\vert =\frac{M_{A}-m_{A}}{\sqrt{m_{A}}%
M_{A}}\text{.}  \label{DtrapLaw}
\end{equation}%
Interestingly, one sees that $|b_{e}^{A}(\infty )|$ is only related with the
number $m_{A}$ of initially excited emitters and the total number $M_{A}$ of
emitters of type $A$, thus still exhibiting a simple trapping law of
emission dynamics. In Fig.~\ref{Figure3} (a), we plot the theoretical time
evolution of $|b_{e}^{A}((t)|$ with different numbers $m_{A}$ of initially
excited QEs for $M_{A}=5$. It is seen that $|b_{e}^{A}((t)|$ asymptotically
approaches $(M_{A}-m_{A})/(\sqrt{m_{A}}M_{A})$. On top of this remarkably
simple behavior, $|b_{e}^{A}((t)|$ is seen to be stabilized or trapped, but
with some small-amplitude oscillation behavior. This oscillation behavior
can be traced back to the contribution from the above-neglected photon-QE
bound states, associated with the second term in Eq.~(\ref{Dtrapping}). One
would imagine that if the coupling strength $V_{A}$ is tuned to be smaller
so that the theoretical condition $(s+i\Omega _{B})m_{A}V_{A}^{2}F(s)/\{%
\sqrt{m_{A}}(is-\Omega _{A})[G_{2}(s)]^{\prime }\}|_{s=\tilde{\varepsilon}%
_{n}}\ll 1$ is better satisfied, then the said oscillations will become less
obvious.

As a final interesting check to motivate future studies, let us now
investigate the emission dynamics via $|b_{e}^{A}((t)|$ in Fig.~\ref{Figure3}
(b) with the condition $V_{B}/(2J)=0.6$. Under this stronger coupling
condition from type-B emitters, the role of the second term in Eq.~(\ref%
{Dtrapping}) or the contributions from the photon-bound states can no longer
be neglected. Indeed, as we see from the actual results, the previously
identified trapping law is completely broken due to the strong interplay
between type-A and type-B QEs.

We conclude this section with more qualitative discussions. Due to the
presence of two different types of identical QEs, there are two types of
degenerate dark states. Dark states due to type-A QEs have energy $%
E=\Omega_{A}$ whereas the other type of dark states has energy $E=\Omega_{B}$%
. However, because of the orthogonality of these two different types of dark
states, only the dark states with $E= \Omega_{A}$ makes a difference to the
emission dynamics if only type-A QEs are initially excited. Nevertheless,
this simple picture is valid only if the impact of population trapping from
the photon-QE bound states is negligible. In the parameter regime where the
population trapping law still persists, it is seen that under the condition $%
m_{A}\ll M_{A}$, the spontaneous emissions from type-A QEs can be greatly
inhibited by the presence of dark states.

\section{Discussion and conclusion}

We have studied the single-photon collective emission dynamics in a
one-dimensional waveguide array system. Assuming that the size of the
ensemble of QEs is much smaller than the wavelength of the radiation field,
we have neglected the spatial difference between the QEs. Our model system
supports stable subradiant states composed of dark states that preserve the
collective excitation of QEs. Unlike the trapping regime caused by the
photon-QE bound states, we find that the long-time emission dynamics of the
subradiant states can be characterized by a unified population trapping law.
This trapping law has nothing to do with the dispersion of the bosonic bath
or the coupling strength between the photon field and the QEs. Instead, it
is only related with the number of initially excited QEs and the total
number of QEs. When more than one type of QEs are present, a similar
trapping law persists if the effect of the composite photon-QE bound state
can be neglected.

Finally, we discuss the possible experimental platform consisting of
transmon qubits and coupled superconducting resonators which have been
realized in recent years \cite%
{Fitzpatrick17,Sundaresan19,Ferreira21,McKay15,Scigliuzzo21,Kim21}. In such
systems, the hopping energy $J\approx 20$-$730(2\pi)$ MHz. The
qubit-resonator coupling strength $V$ is in the range of $5$-$300(2\pi)$
MHz. Thus the key parameter $V/(2J)\ll 1$ can be achieved with the existing
technology. The frequencies of transmon qubits can be controlled in the
range of $1$-$10(2\pi)$ GHz \cite{Blais21,Gu17}, which is similar to the
range of the resonance frequency of each resonator, hence being sufficient to yield
small detuning $\Delta $ or near-resonance conditions considered in this
work.

\bigskip



\end{document}